\documentclass{JAC2000}
\usepackage{graphicx}

\begin{document}
\title{Multi-bunch generation by thermionic gun 
%\thanks{Work supported by long suffering spouses and colleagues}
}

\author{M. Kuriki, H. Hayano, T. Naito, KEK, Tsukuba, Ibaraki, Japan\\
K. Hasegawa, Scientific university of Tokyo, Noda, Chiba, Japan}

\maketitle

\begin{abstract} 

KEK-ATF is studying the low-emittance multi-bunch electron beam for the
future linear collider. In ATF, thermionic gun is used to generate 20
bunches electron beam with the bunch spacing of 2.8 ns. Due to a
distortion of the gun emission and the beam loading effect in the
bunching system, the intensity for each bunch is not uniform by up to 40
\% at the end of the injector. We have developed a system to correct the
gun emission by precisely controlling the cathode voltage with a
function generator. For the beam loading effect, we have introduced RF
amplitude modulation on Sub Harmonic Buncher, SHB. By these technique,
bunch intensity uniformity was improved and beam transmission for later
bunches was recovered from 67\% to 91\%, but intensity for first five
bunches is still lower than others.
\end{abstract}

\section{Introduction}
KEK-ATF is a test facility to develop the low emittance multi-bunch beam
and beam instrumentation technique for the future linear collider.  That
consists from 1.5 GeV S-band linac, a beam transport line, a damping
ring, and a diagnostic extraction line.

In the linac, the electron beam is generated by a thermionic electron
gun. Typical intensity is $\rm 1.0\times 10^{10}$ electron/bunch. The
bunch length is compressed from 1 ns to 10 ps by passing a couple of
sub-harmonic bunchers, a TW buncher, and the first S-band accelerating
structure. This area is called as injector part. After the injector
part, electron energy becomes 80 MeV.

The electron beam is then accelerated up to 1.3 GeV by 8 of the S-band
regular accelerating section.  One section has two accelerating
structures driven by a klystron-modulator. Klystron is Toshiba E3712
generating 80 MW with a pulse duration of $\rm 4.5 \mu s$ RF. A peak
power of 400 MW with a pulse duration of $\rm 1.0\mu s$ is obtained by
SLED cavity and makes a high gradient accelerating field, 30 MeV/m.

20 of bunches separated by 2.8 ns are accelerated by one RF pulse. This
multi-bunch method is one of the key technique in the linear collider.

In April 2000, we achieved horizontal emittance $\rm 1.3\times 10^{-9}
rad.m $, vertical emittance$1.7\times 10^{-11} rad.m $ (both for
$2.0\times 10^9 electron/bunch$, single bunch mode )\cite{emittance}
which are almost our target. 

In November 2000, we have started the multi-bunch operation. The
commissioning was successfully done. Due to lack of the instrumentation
device for the multi-bunch diagnostic, emittance for each bunch is not
measured yet.

\section{Multi-bunch beam generation}\vspace{-.2cm}

The gun assembly consists from a thermionic gun, Grid pulser, and a
high voltage gun pulser.

The thermionic gun, is a triode type, EIMAC Y796. The electron current
is controlled by Grid bias. 

\begin{figure}[htbp]
\includegraphics*[width=3in]{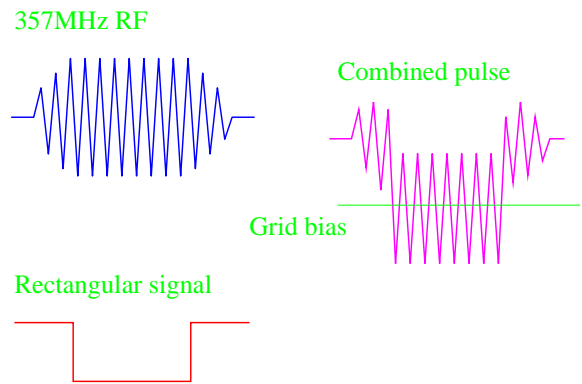}
\vspace{-.3cm} \caption{
To omit the rise and fall edge of 357 MHz RF signal,
a rectangular signal is combined. Grid bias is determined to clip
uniform multi-bunch beam.\label{fig:gunrf}} 
\end{figure}

To make a multi-bunch electron beam with a bunch spacing of 2.8 ns, 357
MHz RF signal is applied to the GUN cathode. 357MHz ECL level RF signal
is amplified by a power amplifier. This output has a pulse height of 400
V peak-to-peak, but the amplitude is gradually changing at the rise and
fall edge as shown in FIG. \ref{fig:gunrf}.

If the RF signal is directly applied to Gun cathode, bunch intensity
becomes not flat. To get uniform bunches, an additional rectangular
pulse is combined as shown in FIG \ref{fig:gunrf}. The grid bias is
determined that the rectangular pulse clips out the flat part of the RF
signal. Finally, only the flat part of the RF pulse is obtained as real
beam current.

\vspace{-.2cm}
\section{Emission correction}
\begin{figure}[htbp]
\centering
\includegraphics*[height=2.5in]{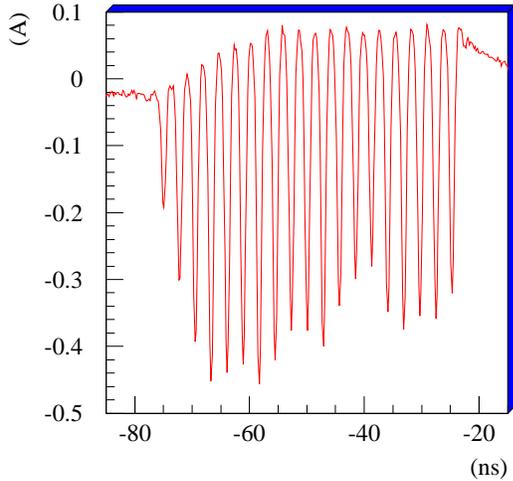}
\vspace{-.5cm}\caption{Multi-bunch beam measured by a current
transformer. Vertical axis shows the beam current in A. Grid bias was
240 V.\label{fig:raw}}
\end{figure}

\vspace{-.2cm}
FIG. \ref{fig:raw} shows the multi-bunch beam generated by thermionic
gun. The vertical and horizontal axes show time in ns and the beam
current in A respectively. The grid bias was set to 240 V. The left side
is early bunch. The beam current is measured by a current transformer
which is set right after the gun exit. The current transformer measures
the beam current as the induction voltage, so the output decays with a
time constant.

Intensity for the first three bunches is still increasing.  This
behavior is due to the rounded rising edge of the clipping rectangular
pulse.

In addition, several bunches around 13th and 14th have lower intensity
than others. A study for the gun emission \cite{linac99} demonstrated
that the gun response to the rectangular pulse reproduced this dip, but
the reason was not fully understood. This is not any problem on the
electrical circuit such as reflection signal because any dip was not
observed in direct measurement of the rectangular pulse applied to the
cathode.

\begin{figure}[htbp]
\includegraphics*[height=2.5in]{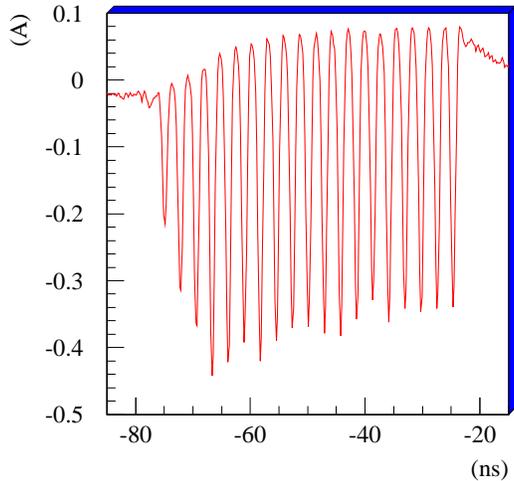}
\vspace{-.5cm}\caption{Multi-bunch beam applying the correction
signal. Grid bias was 240 V. \label{fig:corr}}
\end{figure}
To correct this dip, an additional signal source was introduced. The
correction signal is produced by a function generator, Tektronix AWG 510
which can make an arbitrary waveform with 1 GHz clock speed. The signal
is transfered to the gun high voltage station through an optical cable ,
amplified 20 W RF amplifier, and combined with the main signal through a
resistive power combiner. Typical amplitude of the correction signal is
30 V which is roughly 10\% of the main signal applied to the gun
cathode.

FIG. \ref{fig:corr} shows the gun output by applying the correction
signal.  The grid bias was set to 240 V. The first three bunches have
still current lower than others, but the large dip on 12-15th bunches in
FIG. \ref{fig:raw} was well compensated.

\vspace{-.3cm}
\section{SHB amplitude modulation}\vspace{-.2cm}
Electron beam generated by the thermionic gun has approximately 1 ns
bunch length which is larger than acceptance of S-band acceleration. 
A couple of 357 MHz standing wave Sub-harmonic bunchers, and a traveling
wave S-band buncher are placed to gather electrons into the S-band
acceptance, $\rm 10 - 20 ps$.

In multi-bunch operation, the bunching field is decreased by beam
induced field, i.e. wake field. This is the beam loading effect. Beam
loading effect is larger for later bunch, so the condition becomes worse
for the later bunches.

\begin{figure}[htbp]
\includegraphics*[width=3in]{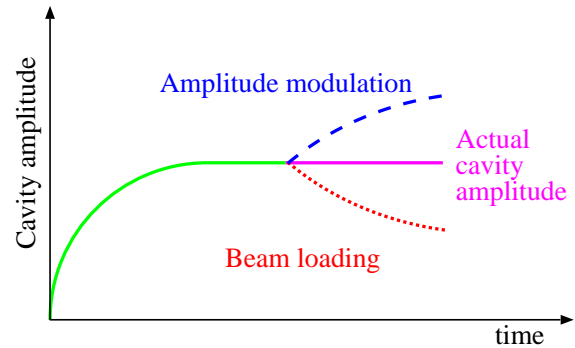}
\vspace{-.4cm}\caption{ In amplitude
modulation method, amplitude of input RF is changed synchronously with
the beam timing. Cavity RF amplitude is then gradually increased with
the filling time as shown by the dashed line. On the other hand, RF
amplitude is decreased by the beam loading effect as shown by the dotted
line. Totally, cavity RF amplitude is kept flat.\label{fig:am}}
\end{figure}

To compensate the beam loading effect, we have introduced amplitude
modulation on pulsed RF for SHBs. In amplitude modulation, the amplitude
of pulsed RF is changed synchronously with the beam
timing. FIG. \ref{fig:am} shows the beam loading compensation by
amplitude modulation schematically. Cavity RF amplitude is then
gradually increased with filling time as shown by the dashed line. On
the other hand, RF amplitude is decreased by the beam loading effect as
shown by the dotted line. Totally, cavity RF amplitude is kept flat. The
bunching quality becomes equal for all bunches.

The beam loading effect always decelerates the followed bunches. In ATF,
the first SHB, SHB1 is operated in deceleration mode and the second,
SHB2 is in acceleration mode. The beam loading effectively increases RF
amplitude in SHB1 and decreases in SHB2 . Modulation sign is then
negative for SHB1 and positive for SHB2.

\begin{figure}[htbp]
\includegraphics*[height=2.6in]{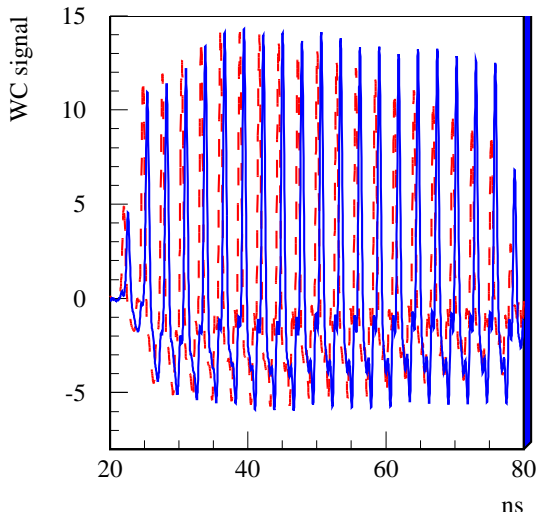}
\vspace{-.4cm}\caption{ Multi-bunch beam
profile by wall current monitor. Horizontal axis shows time in ns.
Vertical axis shows wall current monitor response in V.  The dotted and
solid curves were obtained without and with amplitude modulation on SHB
 RF. \label{fig:wc}}
\end{figure}

Optimization for the amplitude modulation has been done by looking beam
transmission at the end of injector part. A wall current monitor is
placed at the exit of the injector part to observe the beam
current. FIG. \ref{fig:wc} shows the response of the wall current
monitor to the multi-bunch beam. The dotted and solid curves indicate
those obtained with the conventional pulsed RF and the amplitudely
modulated pulsed RF on SHBs respectively. Transmission for the later
bunches was recovered by the amplitude modulation.

The beam loading effect affects the transmission for the later bunches,
then we should investigate the bunch transmission to examine the beam
loading effect.  Since the absolute transmission for each bunch is hard
to measure exactly, the intensity ratio of the early bunch and later
bunch can be used instead of the absolute transmission.

Intensity of the last bunch is much lower than others due to the less
sharpness of the clipping rectangular pulse. Because of that, effect of
the amplitude modulation should be examined by the last second bunch
rather than the last bunch.

The transmission ratio of the second last bunch was 0.67 for the
conventional SHB RF and 0.91 for the amplitudely modulated SHB RF
respectively. 
The most intense bunch was used as the
reference. Improvement of the transmission by the amplitude modulation
was 24\%. 

\begin{figure}[htbp]
\centering
\includegraphics*[height=2.4in]{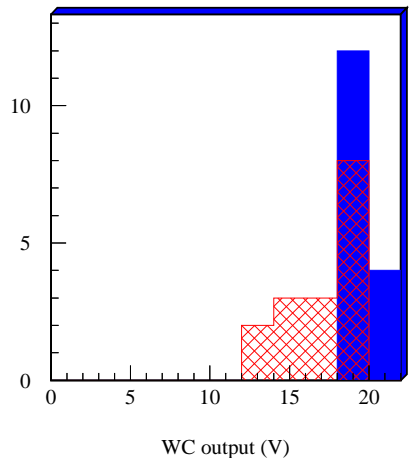}
\vspace{-.4cm}\caption{The horizontal axis shows bunch intensity measured by wall current
monitor. The vertical axis is number of bunches per 1.0 V. The solid and
hatched histograms show those with the amplitude modulation and the
conventional RF on SHB respectively. \label{fig:intensity}}
\end{figure}

FIG. \ref{fig:intensity} shows distributions of bunch intensity for 6th
and later bunches. The peak voltage of wall current monitor is here used
instead of the real beam current. The solid and hatched histograms are
those with the amplitude modulation and the conventional pulsed RF on
SHB respectively. With the amplitude modulation, most bunches are
distributed more than 18 V, but with the conventional RF, bunches are
spread widely from 12 V to 20 V. The amplitude modulation improved the
flatness of intensity for these later bunches.

FIG. \ref{fig:intensity} does not include the first five bunches. The
lower intensity of these bunches is due to the rounded rising edge of
the clipping pulse. That will be one of the main issue on the
multi-bunch operation.

\vspace{-.5cm}
\section{summary}\vspace{-.2cm}
In KEK-ATF, multi-bunch beam was successfully generated by a thermionic
electron gun with bunch spacing of 2.8 ns. The beam already reached to
the extraction line, but the emittance was not measured yet due to lack
of the instrumentation for multi-bunch beam.

Intensity for each bunch is not uniform because of ; 1) gun emission
un-uniformity; 2) beam loading effect. 

For gun emission problem, we have applied a correction signal generated
by an arbitrary function generator to Gun cathode. Bunch intensity
flatness was significantly improved by this emission
correction. However, Gun emission for first five bunches is still lower
than others. That will be one of the main issue in future. 

For beam loading effect, we have introduced amplitude modulation on SHB
RF. The amplitude modulation compensated the beam loading effect and
recovered the beam transmission from 67\% to 91\%.

\end{document}